\documentclass[preprint]{aastex}
\usepackage{amssymb,amsmath,graphicx,endfloat,natbib}

\begin{document}

\title{MHD Effects on Pulsed YSO Jets. I. 2.5-D Simulations}
\author{E. C. Hansen and A. Frank}
\affil{Department of Physics and Astronomy, University of Rochester, Rochester, NY 14627-0171, USA}
\and\author{P. Hartigan}
\affil{Department of Physics and Astronomy, Rice University, 6100 S. Main, Houston, TX 77521-1892, USA}

\begin{abstract}
In this paper we explore the dynamics of radiative axisymmetric MHD jets at high resolution using AMR methods.
The goal of the study is to determine both the dynamics and emission properties of such jets.
To that end we have implemented microphysics enabling us to produce synthetic maps of H$\alpha$ and [S II].
The jets are pulsed either sinusoidally or randomly via a time-dependent ejection velocity which leads to a complicated structure of internal shocks and rarefactions as has been seen in previous simulations.
The high resolution of our simulations allows us to explore in great detail the effect of pinch forces (due to the jet's toroidal magnetic field) within the ``working surfaces'' where pulses interact.
We map the strong H$\alpha$ emission marking shock fronts and the strong [S II] emission inside cooling regions behind shocks as observed with high-resolution images of jets.
We find that pinch forces in the stronger field cases produce additional emission regions along the axis as compared with purely hydrodynamic runs. These simulations are a first step to understanding the full 3-D emission properties of radiative MHD jets.
\end{abstract}


\section{Introduction}
\label{sec:intro}
Hypersonic jets are a common feature of the star formation process.
These narrow, highly-collimated beams of atomic and/or molecular gas have velocities of order $v \sim 100-1000$ km/s.
Jets are thought to form via magneto-hydrodynamic (MHD) processes in the rotating star-disk system.
Less collimated, more massive molecular outflows are also common and these are believed to consist of ambient gas swept-up by jet bowshocks.

Since jets are accretion-driven, outflow properties that change with distance from the source provide important constraints on past temporal variations in the ejection/accretion system over a huge range of timescales from $<5$ to $10^5$ years.
Observations show that jet beams exhibit a series of closely spaced inner ``knots'' within 0.1pc of the source (see \citet{Hartigan11}).
Jet beams also show more distant, well separated, and larger bows or ``bullets'' which in most cases have a clear correspondence in each lobe of the bipolar jet \citep{Zinnecker98}.
The knots and bows seen in atomic jets have line ratios characteristic of internal shock waves.
Thus, significant variations in speed or ejection angles are also required for their formation.
Evidence points to these shocks being caused by supersonic velocity jumps where fast material catches up with slower ejecta \citep{Raga02ApJ,Hartigan05}. 

The natural explanation for velocity variations is via variability in the ejection speed \citep{Raga90} at the jet source.
This scenario finds support in numerical simulations of objects such as HH34 \citep{Raga12}.
The knot/bow spacing and velocity patterns in Class I atomic jets then suggest that multiple modes of velocity variability may be present \citep{Raga02AA}.
An alternative, though related explanation, assumes sub-radial structure in the jet beam \citep{Yirak09,Yirak10}.
These clumps may be an inherent part of the key launching process, or they may form close to the source via instabilities.
It has been shown \citep{Yirak09} that random clumps with a distribution of velocities can mimic periodic pulsations at any given epoch.

The acquisition of multiple-epoch emission line images from HST adds another degree of complexity to this picture.
These observations now span enough time (10 years) to reveal how the knots evolve.
Images of the classic large-scale bowshocks in HH1\&2, HH~34, and HH~47 \citep{Hartigan11} show evidence for a variety of phenomena related to jet propagation including standing deflection shocks where the precessing jet beam encounters the edges of a cavity, and a strong bow shock encounters a dense obstacle on one side.
Knots along the jet may brighten suddenly as denser material flows into the shock front and fade on the cooling timescale of decades.
Multiple bow shocks along working surfaces sometimes overlap to generate bright spots where they intersect, and the morphologies of Mach disks range from well-defined, almost planar shocks to small reverse bow shocks as the jet wraps around a denser clump.
The bow shocks themselves exhibit strong shear on the side where they encounter slower material, and they show evidence for Kelvin-Helmholtz instabilities along the wings of the bows. 

The wealth of detail in these multi-epoch observations requires high resolution simulations that can track both evolution of the knots (and their substructure) as well as their emission properties.
However, carrying forward such simulations is difficult because an accurate treatment of post-shock micro-physics (ionization and chemistry) requires that shock cooling zones be resolved.
Since the cooling zones can be of order $10^{-4}$ of the jet radii, only highly parrallelized AMR simulations can be expected to capture both the global and local dynamics of radiative pulsed MHD jets.
The goal of this study is to take the first step by exploring pulsed jet evolution with magnetic fields in axisymmetry (2.5-D). 

Our study focuses on the pulsed MHD jet dynamics and seeks to produce synthetic observations in both H$\alpha$ and [S II] (6716 \AA ~+ 6731 \AA ~lines).
H$\alpha$ typically marks shock fronts, and [S II] peaks in cooling regions behind shocks (\citet{Heathcote96}).
Both steady state and time-dependent models show that shock fronts and cooling regions are affected by magnetic fields.
In cases where the magnetic field is perpendicular to the shock normal, the magnetic pressure from the field can inhibit radiative cooling by keeping the density from increasing behind the shock and extending the cooling region where [S II] emission is expected. 
	
Previous numerical studies have explored in great detail the propagation of supersonic jets (e.g. \citet{Norman82,Stone93}).
Radiative processes have been shown to affect the morphology of the jet head, beam, cocoon, as well as the internal working surfaces (\citet{Hartigan87,deColle08}).
Numerical simulations have also shown that magnetic fields can create pinch forces which enhance bow shock velocities and affect the jet beam evolution (\citet{Frank98}).
The production of synthetic emission maps in this context adds a level of complexity to the analysis of such jets, and it is less common in numerical studies.
Emission due to variable velocity jet models has been studied by several groups (e.g. \citet{Kajdic06,deColle06}).
In this study we build on past numerical simulations (as well as observations) by combining radiative losses and magnetic fields in simulations at high enough resolution to capture the cooling zone and with a larger set emission lines. 
	
The simulations presented in this work use a similar initial set-up as  \citet{deColle06}.
Most studies of this type have focused on the morphology of the jet, but de Colle and Raga extended the analysis to predicting the H$\alpha$ emission.
We have further extended their work by running our own 2.5-D simulations at a higher resolution and by producing syntehic emission maps of the [S II] lines as well.
The [S II] lines have been studied in numerical simulations as early as the work done by \cite{Raga91}.
Detailed 1-D simulations of a variable wind, which produced forward and reverse shocks, were conducted by \citet{Hartigan93}, and they focused on the resulting [S II] lines.
Line ratios have been used to determine ionization fractions and shock velocities \citep{Hartigan94}.
Other works that have studied the [S II] lines from simulations of jets include \citet{Rubini07} and \citet{Tesileanu12}.
The need for higher resolution in terms of dynamics alone has been shown by \citet{Yirak10} who demonstrated the lack of convergence in radiative shock-clump simulations unless the cooling zone is resolved with at least $\sim 10$ zones.
The production of accurate synthetic observations also requires resolution of the cooling zone.
Thus our work in 2.5-D, where such high resolutions are feasible, helps to articulate classes of behavior which may be more difficult to resolve in 3-D studies.

The structure of our paper is as follows: in Section~\ref{sec:sims}, we present our numerical methods as well as our initial conditions.
We then include a brief discussion on resolution.
Section~\ref{sec:theory} contains a numerical analysis of the forces within the internal working surfaces.
We present and explain our results in Section~\ref{sec:results}.
Section~\ref{sec:random} we explore the morphological consequences of a randomly pulsed jet model.
Finally, in Section~\ref{sec:discconc} we present some more discussion and our conclusions. 

\section{Numerical Simulations}
\label{sec:sims}
\subsection{Methods}
\label{subsec:methods}
The simulations were carried out using AstroBEAR, a highly parrallelized adaptive mesh refinement (AMR) multi-physics code.
See \citet{Cunningham09,Carroll12} for a detailed explanation of how AMR and MHD are implemented.
More details of the code can also be found at http://bearclaw.pas.rochester.edu/trac/astrobear.
Here we provide a brief overview of the physics implemented for this study.
The code solves the equations of ideal MHD with non-equilibrium cooling:

\begin{subequations}\label{group1}
\begin{gather}
\frac{\partial \rho}{\partial t} + \boldsymbol{\nabla} \cdot \rho \boldsymbol{v} = 0 ,\ \label{1a}\\[\jot]
\frac{\partial \rho \boldsymbol{v}}{\partial t} + \boldsymbol{\nabla} \cdot (\rho \boldsymbol{v} \boldsymbol{v} + P_{tot}\boldsymbol{I} - \boldsymbol{B}\boldsymbol{B}) = 0 ,\ \label{1b}\\[\jot]
\frac{\partial E}{\partial t} + \boldsymbol{\nabla} \cdot ((E + P_{tot}) \boldsymbol{v} - (\boldsymbol{v} \cdot \boldsymbol{B}) \boldsymbol{B}) = - L ,\ \label{1c}\\[\jot]
\frac{\partial \boldsymbol{B}}{\partial t} + \boldsymbol{\nabla} \cdot (\boldsymbol{v} \boldsymbol{B} - \boldsymbol{B} \boldsymbol{v}) = 0 ,\ \label{1d}\\[\jot]
\frac{\partial n_i}{\partial t} + \boldsymbol{\nabla} \cdot n_i \boldsymbol{v} = C_i ,\ \label{1e}
\end{gather}
\end{subequations}
where $\rho$ is the mass density, $\boldsymbol{v}$ is the velocity, $P_{tot}$ is the total pressure (thermal + magnetic) defined as $P_{tot} = P_{gas} + \frac{1}{2} B^2$, $\boldsymbol{I}$ is the identity matrix, $\boldsymbol{B}$ is the magnetic field normalized by $\sqrt{4 \pi}$, and $E$ is the total energy such that $E = \frac{1}{\gamma - 1} P_{gas} + \frac{1}{2}\rho v^2 + \frac{1}{2} B^2$ (with $\gamma = \frac{5}{3}$ for an ideal gas).
$L$ is the cooling source term which will be a function of number density, Temperature and ionization. $n_i$ is the number density of species $i$, and $C_i$ is the sum of the ionization and recombination processes for species $i$.

The equations above represent the conservation of mass \eqref{1a}, momentum \eqref{1b}, energy \eqref{1c}, and magnetic flux \eqref{1d}.
Equation \eqref{1e} represents the evolution of the number densities of the different atomic species.

The cooling source term on the right hand side of equation \eqref{1c} is implemented in AstroBEAR through calculations using the ionization rate equations.
The rates are then used to determine the ionization and recombination energy losses from both hydrogen and helium.
These non-equilibrium calculations are always done regardless of the temperature.
Cooling from metal excitation is calculated from one of two different tables depending on the temperature. 
Below $10^4 K$, we use a table constructed by Hartigan which uses strong charge exchange cross sections to lock the ratios of NII/NI and OII/OI to HII/HI, and it solves the multilevel atom to derive a volume emission cooling term.
Above $10^4 K$, we use a modified version of the Dalgarno \& McCray cooling curve (\citet{Dalgarno72}).
It is modified by subtracting the contributions from H and He, leaving only the metal components.
A more detailed description of Hartigan's table and other cooling processes relevant to radiative shocks will be given in a future paper (\citet[in preparation]{Hartigan15}).

Using equation \eqref{1e} allows us to keep track of the number densities of the neutral and ionized species.
There is a total of 8 species tracked in the code: neutral and ionized hydrogen, neutral, ionized, and doubly ionized helium, and SII, SIII, and SIV.
Tracking the hydrogen and helium species allows us to track their ionization fractions and thus the electron number density of the gas which is required for generating synthetic emission maps.
Below $10^4 K$, all S is assumed to be SII, and above $10^4 K$, the ionization and recombination rates are used to track the amount of SIII and SIV.
This was employed to more accurately track the ionization state of S and hence produce more accurate [S II] maps.

\subsection{Initial Conditions}
\label{subsec:init}
Our simulations occur in a domain size of $(L_r, L_z)= (3 \times 10^{16}, 3 \times 10^{17})$ cm.
The base resolution is $42 \times 420$ with 8 additional levels of AMR which corresponds to an effective resolution of about $2.79 \times 10^{12}$ cm/cell.
This is equivalent to 1792 cells per jet radii.
We use a reflective boundary conditions at $r = 0$.
At $z = 0$ the boundary is overwritten by the jet ejection conditions.
The outer boundaries use open conditions.
	
The jet has an initial radius $R_j = 5 \times 10^{15}$ cm and propagates in the positive z-direction.
The initial and mean ejection velocity of the jet is $v_0 = 200$ km/s.
Clumps, or knots, along the jet axis are created by pulsing the jet with a time-varying ejection velocity.
For simplicity, we use a sinusoidal velocity perturbation of the form

\begin{equation}
v(t) = v_0 \left( 1 + A \sin \frac{2 \pi t}{\tau} \right) ,\label{2}
\end{equation}
where the amplitude $A = 0.25$, and the period $\tau = 50$ years.
We let the jet propagate for 500 years.

The injected jet density is held constant at 500 cm\textsuperscript{-3}, and the uniform ambient density is 100 cm\textsuperscript{-3}.
The ambient temperature is set at 10,000 K.
We impose hydromagnetic pressure equilibrium inside the jet using a prescription which is very similar to the one used by \citet{Lind89} (this profile has been used by many others (e.g. \citet{Frank98,OSullivan00}).
Thus, the initial toroidal magnetic field profile takes the form

\begin{equation}
B(r) = \left\{
\begin{array}{ll}
B_m \frac{r}{R_m} & 0 \leq r < R_m \\
B_m \frac{R_m}{r} & R_m \leq r < R_j \\
0 & R_j \leq r.
\end{array}
\right. \label{3}
\end{equation}

This is also similar to the field set up used by \citet{Stone00}.
In equation \eqref{3}, $B_m$ is a free parameter defined as the magnetic field strength at the radius $R_m$.
$R_m$ is a free parameter, and we have chosen it to be $0.6R_j$.
To inject the field in the jet we use the vector potential defined in the "ghost zones" of the grid.
We define a vector potential as

\begin{equation}
\boldsymbol{A}(r) = \left\{
\begin{array}{ll}
- \frac{1}{2} B_m \frac{r^2}{R_m} \boldsymbol{\hat{v}} & 0 \leq r < R_m \\
- B_m R_m \left( \log \left( \frac{r}{R_m} \right) + \frac{1}{2} \right) \boldsymbol{\hat{v}} & R_m \leq r < R_j \\
- B_m R_m \left( \log \left( \frac{R_j}{R_m} \right) + \frac{1}{2} \right) \boldsymbol{\hat{v}} & R_j \leq r.
\end{array}
\right. \label{4}
\end{equation}

The vector potential $\boldsymbol{A}$ is in the direction of jet propagation which is denoted by the unit vector $\boldsymbol{\hat{v}}$.
Thus, by defining our magnetic field as the curl of $\boldsymbol{A}$, we ensure that our field is divergence free.
Furthermore, our resulting field has the same profile as equation \eqref{3}.

Finally, to satisfy the condition of hydromagnetic equilibrium, the pressure is initialized with

\begin{equation}
P(r) = \left\{
\begin{array}{ll}
\left( \alpha + \frac{2}{\beta_m} \left(1 - \frac{r^2}{R_m^2}\right)\right) P_{amb} & 0 \leq r < R_m \\
\alpha P_{amb} & R_m \leq r < R_j \\
P_{amb} & R_j \leq r,
\end{array}
\right. \label{5}
\end{equation}

where $P_{amb}$ is the ambient pressure and $\alpha = 1 - \frac{1}{\beta_m}\frac{R_m^2}{R_j^2}$.
	
We have run 4 different models with varying magnetic field strengths.
The models have $\beta_m = \infty$, 5, 1, and 0.4 (where $\beta_m = \frac{2 P_{amb}}{B_m^2}$).
These values correspond to no field (hydro), a weak, intermediate field, and strong field respectively.
For simplicity, $\beta_m$ will henceforth be referred to as $\beta$.

\subsection{Resolution}
\label{subsec:res}
As noted in the introduction, resolution is a key issue for numerical simulations of radiative jets.
Physical processes associated with non-equilibrium cooling determine when a simulation can be considered well resolved.
The principle requirement is to resolve the cooling length $L_{cool}$.
As shown by \citet{Yirak10}, failure to resolve the cooling length means that behavior such as the development of certain instabilities will be lost.
One can still capture instabilities such as the pinch mode instability without resolving the cooling length, but cooling instabilities and instabilities indirectly caused by strong cooling require sufficient cooling length resolution.
\citet{Yirak10} found that at least 10 zones per $L_{cool}$ were required to capture these processes.
Creating accurate synthetic maps of emission lines like [S II], which occurs in the cooling zones, also requires this resolution.  

In this work, we have defined $L_{cool}$ as the distance required for shocked material to cool down to 1000 K.
Stronger shocks will cool faster and thus have shorter cooling distances.
Therefore, it is necessary to consider the strongest shock in a given simulation in order to determine the minimum cooling length that needs to be resolved.

The head of the jet creates a very strong bow shock which we do not attempt to fully resolve.
We are only concerned with the internal shocks located at the internal working surfaces.
These shocks are much weaker because they are at lower relative velocities.
The strongest internal shock should occur when the jet velocity is at a maximum and is overrunning the slower injected material.
At this moment, a pair of forward and reverse shocks is generated.
The velocity of each of these shocks will be related to the injected sinusoidal velocity as $v_{max \, shock} = v_oA$ (see equation \eqref{2}).
Thus, the total velocity gradient across the shock pair is $2v_oA$.

Therefore, the strongest internal shocks in our simulaions would have a velocity of 50 km/s.
That is the peak jet velocity of 250 km/s running into slower material that is at 200 km/s (reverse shock), or the 200 km/s material running into 150 km/s material (forward shock).
AstroBEAR was used to run a 1-D radiative shock model with the aforementioned shock velocity and jet parameters which yielded a cooling length of approximately 2.50 AU.
This calculation was only done for the $\beta = \infty$ case.
Since the presence of a magnetic field would only increase $L_{cool}$ we can consider this to be the smallest cooling zone length that needs to be resolved.

\begin{deluxetable}{ccccccccc}

\rotate


\tablecolumns{9}
\tablewidth{0pt}

\tablecaption{Resolutions From Various Jet Simulation Papers}

\tablenum{1}
\label{table:res}

\tablehead{\colhead{Paper} & \colhead{MHD} & \colhead{Lines} & \colhead{$(L_r, L_z)$} & \colhead{$R_j$} & \colhead{$L_{cool}$} & \colhead{Cells/AU} & \colhead{Cells/$R_j$} & \colhead{Cells/$L_{cool}$} \\
\colhead{} & \colhead{} & \colhead{} & \colhead{(AU)} & \colhead{(AU)} & \colhead{(AU)} & \colhead{} & \colhead{} & \colhead{}}

\startdata
This Paper & yes & H$\alpha$, [S II] & (2005, 20054) & 334 & 2.50 & 5.36 & 1792 & 13.40 \\
\citet{deColle06} & yes & H$\alpha$ & (2005, 20054) & 334 & 2.50 & 0.09 & 30 & 0.22 \\
\citet{Raga07} & no & H$\alpha$, [O I] & (1671, 6685) & 134 & 4.16 & 9.80 & 1314 & 40.77 \\
\citet{Tesileanu12} & yes & [S II], [O I], [N II] & (400, 1200) & 20 & 0.16 & 40.96 & 819.2 & 6.55 \\
\enddata


\end{deluxetable}

We have compared this cooling length and resolution with other papers, and we report our findings in Table~\ref{table:res}.
Using parameters from \citet{Tesileanu12}, the AstroBEAR 1-D model results in a cooling length of 0.16 AU.
This is mainly due to the increased density.
This is for their models which have $n_{jet} = 10^4$ cm\textsuperscript{-3}.
They also have models at $n_{jet} = 5$ x $10^4$ cm\textsuperscript{-3}, which would yield an even smaller minimum cooling length.
Note that although our physical resolution is lower than \citet{Tesileanu12}, our cooling length resolution is actually higher.

The same calculation was also done with parameters from \citet{Raga07}.
The resulting cooling length was 4.16 AU.
Their simulations also have a higher density, but a smaller maximum internal shock velocity of 40 km/s.
This is smaller than the cooling length that they reported and used which was 10 AU.
This is likely due to the fact that AstroBEAR has more realistic and thus stronger cooling down to 1000 K.
Note that while they resolve the cooling length with more zones their work was purely hydrodynamic and did not contain magnetic fields.

\section{Analytic Models of MHD Radiative Working Surfaces}
\label{sec:theory}
We begin with an exploration of an analytic model of the forces acting on the internal working surfaces with magnetic fields.
We focus on the forces driving lateral motions ($v_r$) in the post-shock gas.
Our analysis extends that conducted by \citet{deColle08}.

The magnetic pinch force $F_B$ and the gas pressure force $F_P$ are both at work in the region between the pair of MHD shocks that constitutes the magnetized working surface.
The equations describing these forces are

\begin{equation}
F_B = -\frac{B(r)}{4 \pi r} \frac{d}{dr}(rB(r)) ,\label{6}
\end{equation}

\begin{equation}
F_P = -\frac{d}{dr}P(r) .\label{7}
\end{equation}

Hence, the net force would be $F = F_B + F_P$.
If $F$ is negative, then the ``clump'' or ``knot'', which comprises the gas between the shocks, will be driven towards the jet axis ($v_r < 0$).
If $F$ is positive, then the clump expands away from axis.

Substituting the magnetic field and pressure profiles from our MHD equilibrium equations \eqref{3} and \eqref{5} into equations \eqref{6} and \eqref{7} recovers the fact that before gas encounters the shocks both forces go to zero for $r > R_m$.
For $r < R_m$, the forces are balanced.
Application of the shock jump conditions allows us to see how this balance is disrupted.
Note that due to the radial dependence of the pressure, the sound speed is radially dependent.
Thus the Mach number characterizing the strength of the shock is also radially dependent.
For $r > R_m$, the pressure is constant, so the Mach number is constant.
This means that the post-shock gas pressure and magnetic pressure will still be a constant in this region.
Therefore, we only need to concern ourselves with the region where $r < R_m$.

After applying shock jump conditions, the net force for $r < R_m$ can be written as

\begin{equation}
F(r) = (\chi(r) - \eta(r)^2) \frac{B_m}{2 \pi R_m} B(r) ,\label{8}
\end{equation}
where we have defined $\chi(r)$ as the pressure jump and $\eta(r)$ as the density (or magnetic field) jump.
In the weak shock limit $\chi(r)$ and $\eta(r)$ approach one, and the radial force goes to zero.
For strong shocks radiative cooling is expected to dominate the post-shock flows allowing us to consider isothermal shock jump conditions ($\gamma \sim 1$).
In this limit the pressure jump and density jump are equal, $\chi(r) = \eta(r)$.
Thus, equation \eqref{8} becomes

\begin{equation}
F(r) = (\eta(r) - \eta(r)^2) \frac{B_m}{2 \pi R_m} B(r) .\label{9}
\end{equation}

Thus in the isothermal limit the radial force is always negative and the clump is driven towards the axis via compression.
This is consistent with \citet{deColle08} who found that the non-radiative cases were dominated by radial expansion driven by thermal pressure forces, and the radiative cases were dominated by magnetic forces with $F_r < 0$.

\begin{figure}
\plotone{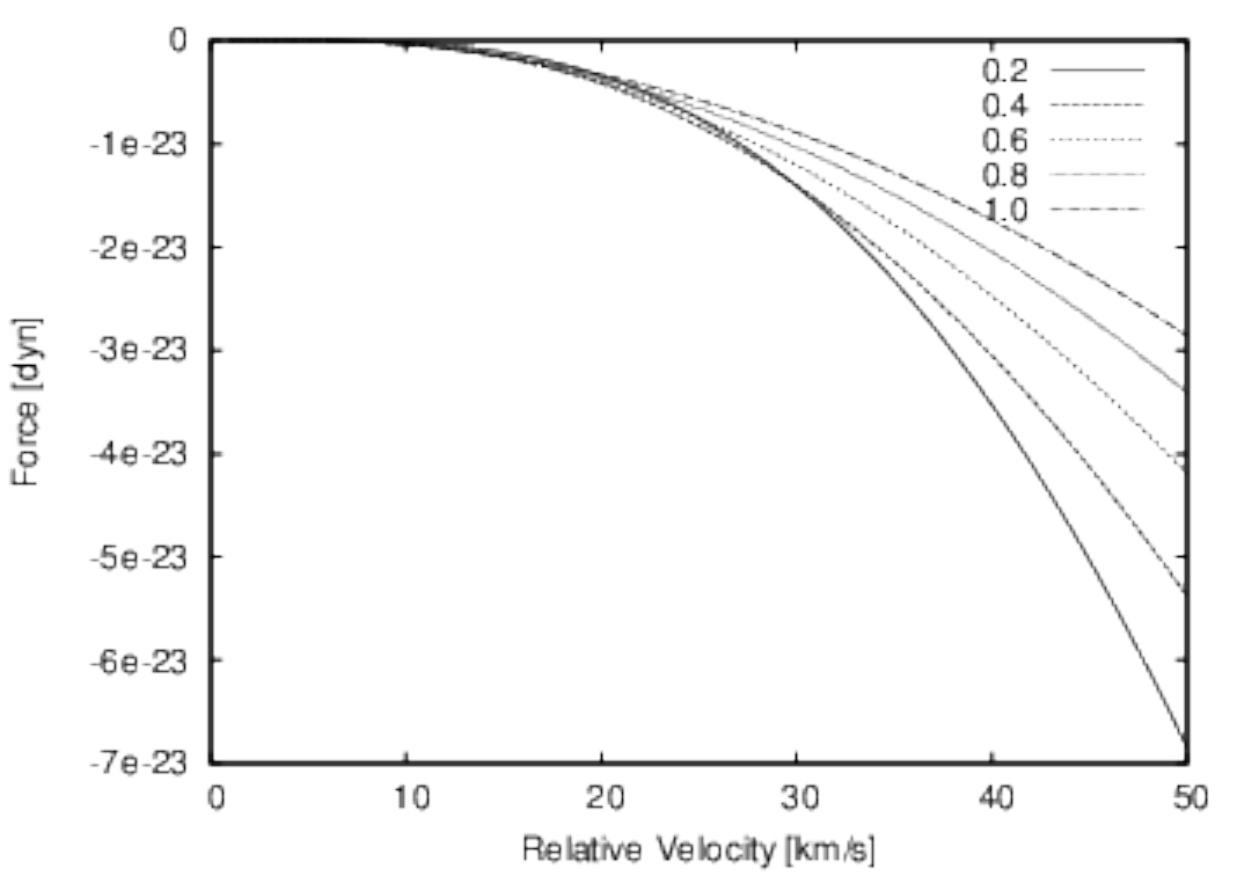}
\caption{The net force (dyn) versus the relative velocity ($v_{jet} - v_{clump}$ in km/s) for the $\beta = 1$ model.
The different lines correspond to different radii (in units of $R_m$). }
\label{fig:forcevel}
\end{figure}

Solving these equations numerically allows us to see both radial dependence of the force and its dependence on the shock strength.
Figure~\ref{fig:forcevel} shows the net force versus relative velocity, $v_{rel}$, which, in this case, is equivalent to the shock velocity.

Note that the magnitude of the inward directed force increases as the shock strength (as given by the Mach number $M=v_{rel}/c$).
This can be understood in terms of the relative amplification of the magnetic field over the pressure force.
Since we are in the isothermal limit, heat generated by the flow is quickly radiated away, however the kinetic energy channeled into $B_\phi$ remains.
Thus increasing shock strength increases the tension force due to the field.
Comparison of the net force at different positions in the jet demonstrates that we should expect an ``inside-out'' compression of the clump between the working surface shocks.
For a given value of the shock strength $M$, there is a a positive gradient in the net force such that $\nabla F_r < 0$.
For constant density, which is the initial condition in our simulations, the outer regions of the jet (out to $R_m$) will experience weaker inward accelerations. 

Note also that based on the results above we do not expect the entire region between the working surface shocks to become compressed towards the axis but only those regions with $F_r < 0$ which implies a division in the axial flow located close to $R=R_m$.

\section{Simulation Results}
\label{sec:results}
The evolution of structure in our jets is driven by the combined effect of velocity perturbations which produce shocks, radiative cooling behind these shocks, and magnetic field pressure and tension.
As discussed above, each pulse steepens into a pair of internal shocks, one facing upstream and one facing downstream \citep{Wilson84}.
These shock pairs travel downstream through the jet beam and continue to evolve based on the strength of cooling and magnetic field effects.
In our jets the cooling parameter (defined as $\alpha = t_{cool}/t_{hydro}$ is such that $\alpha << 1$.
Thus we expect that thermal energy gained as the high speed jet flow passes through the shocks is quickly radiated away producing a cold dense slab.
This slab will be susceptible to a variety of instabilities such as RT and non-linear thin shell modes \citep{Blondin96}.
Our models will not show all of these instabilities because our simulations employ axisymmetry (2.5-D).
The effect of magnetic fields will primarily manifest through pinch forces which will pull material towards the axis.
Once each pulse steepens into a shock, the analysis of Section~\ref{sec:theory} will hold.
Thus we expect to see higher densities along the axis, between the working surface shock-pairs, as the pulses propagate through the jet beam.
Recall that the magnetic force (equation \eqref{6}) can be decomposed into two components,

\begin{equation}
F_B = F_h+F_{bp} = (\frac{1}{c}) (\vec{J}\times\vec{B})=(\frac{B^2}{4 \pi R})\hat{\bf n}+\vec{\nabla}_\perp\frac{B^2}{8\pi}) ,\label{10}
\end{equation}
where $F_h$ is the ``hoop stress'' directed towards the local center of curvature of a field line, and $F_{bp}$ is the magnetic pressure that is directed perpendicular to local gradients.

\begin{figure}
\plotone{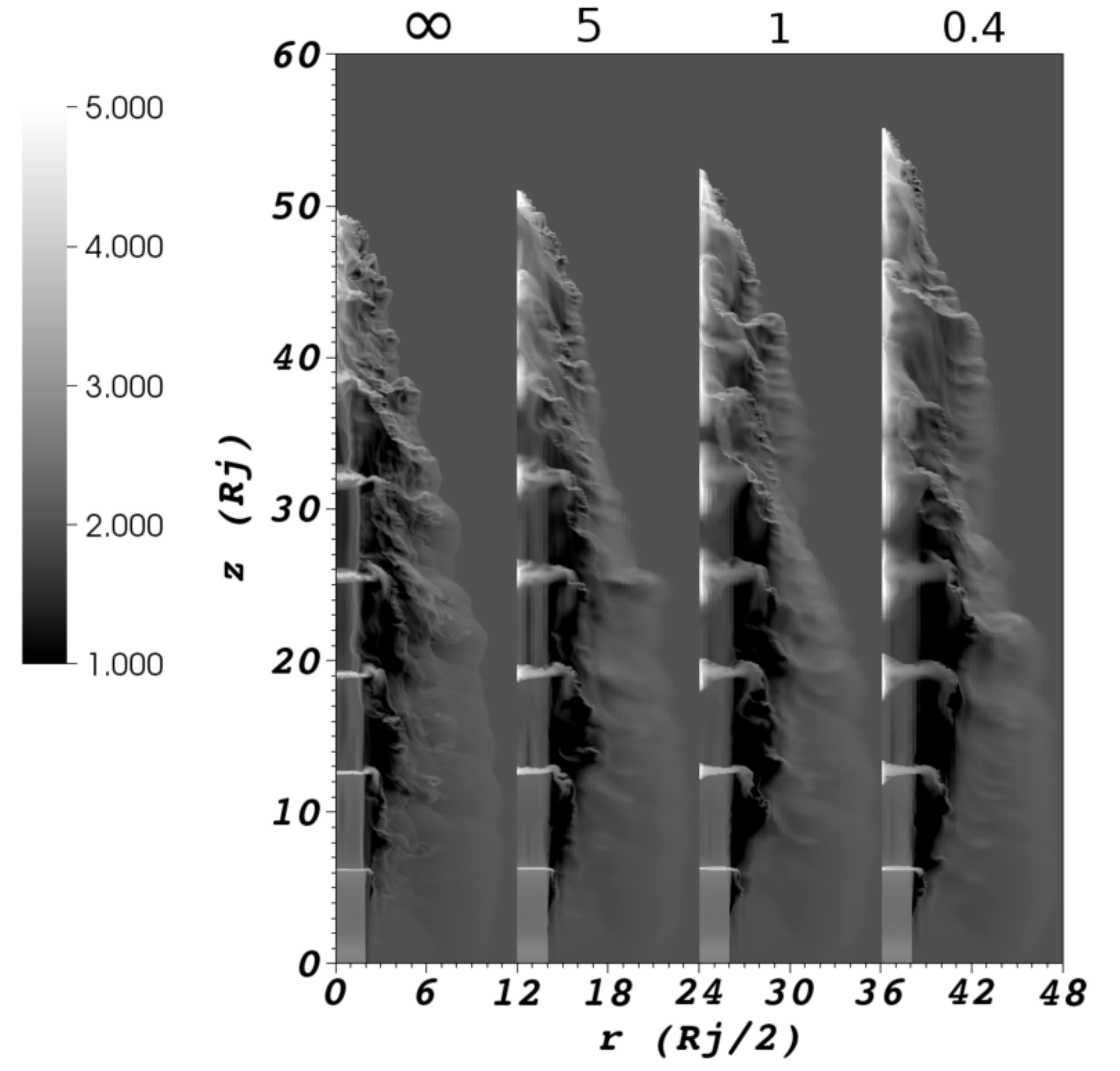}
\caption{Effect of magnetic field on jet axial density.
Number density (cm\textsuperscript{-3}) for all 4 models: $\beta = \infty, 5, 1, 0.4$ from \emph{left} to \emph{right} respectively at time $t = 500$ yrs.
The left bar shows the logarithm of the number density.
The length scale is such that each tick represents one $R_j$.}
\label{fig:rho}
\end{figure}

Figure~\ref{fig:rho} presents a density map of all four models and demonstrates the global effects of pulsation, fields, and cooling.
The runs shown are with initial magnetic field $B_{\phi,0}$ increasing from left to right: $\beta = \infty; 5; 1; 0.4$. 

We first consider the shape of the jet head.
Previous studies have shown that the effect of torodial fields on the leading shock can drive the creation of stream-lined nose-cones.
These form as shocked gas which would be ejected laterally in the absence of fields, but is forced by the hoop stresses, $F_h$, to remain close to the jet beam.
While such nose-cones have been seen in both adiabatic (e.g. \citet{Clarke86,Kossl90a}) and radiative \citep{Frank98} jet simulations their presence is not universal \citep{Cerqueira01} even in 2.5-D simulations \citep{Kossl90a,Kossl90b}.
When such nose-cones are present in magnetized runs, the propagation speed of the leading bow shock is observed to be higher than equivalent hydrodynamic runs \citep{deColle06}, and the streamlining of the jet head by the hoop stresses has been targeted as the cause.
This is precisely what we see in our high resolution simulations; as magnetic field strength increases, the jet head velocity is higher and the jet propagates farther.
The simulations show this trend even in the presence of strong cooling in which instabilities at the jet head may fragment the shock there.
Finally, we note that nose-cones may not form at all in 3-D runs where non-axial instabilities such as the kink mode are likely to destroy the features \citep{Mignone10}.

\begin{figure}
\plotone{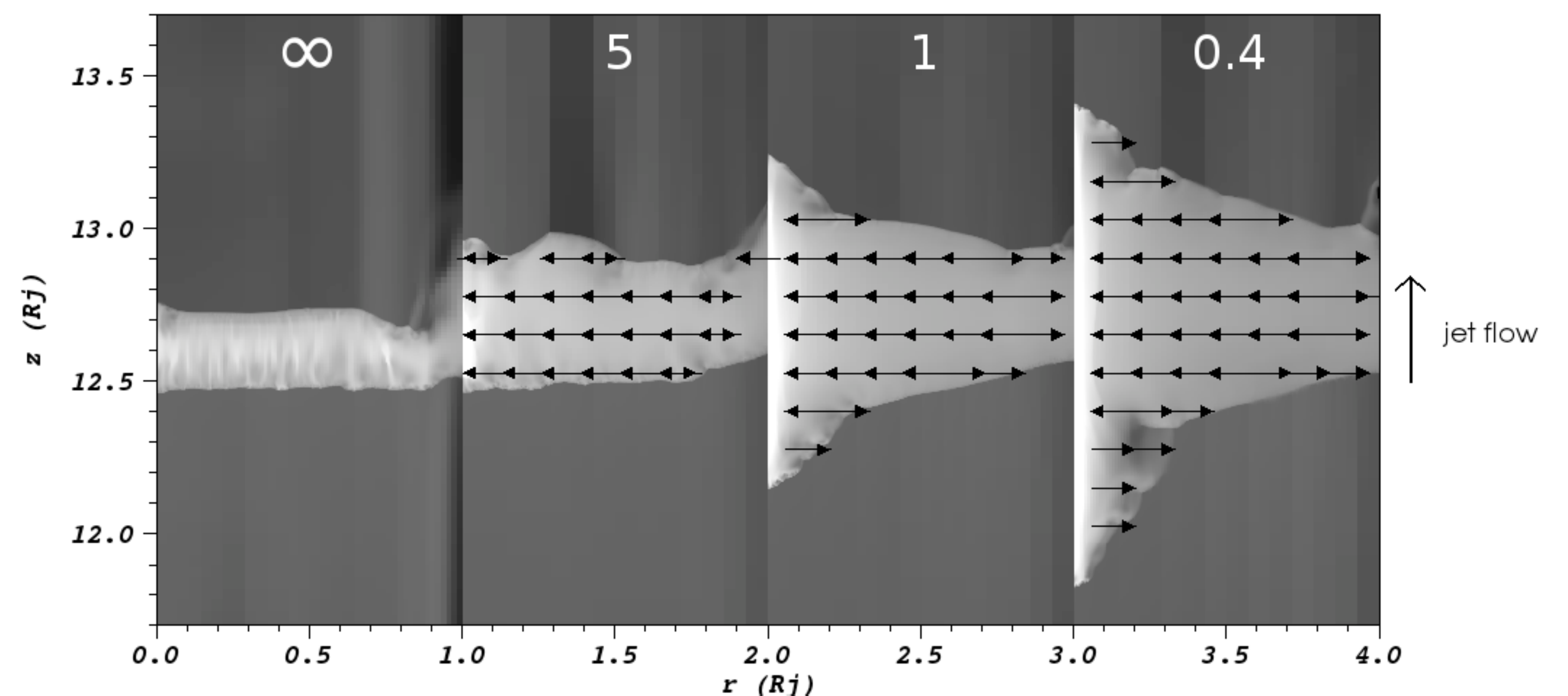}
\caption{Hoop stresses within internal clumps.
An enlarged image of the clump located near $z = 13 R_j$ at $t = 500$ yrs.
The radial domain of each image is from $r = 0$ to $R_j$.
Plot shows number density (cm\textsuperscript{-3}) for all 4 models: $\beta = \infty, 5, 1, 0.4$ from \emph{left} to \emph{right} respectively.
The density scale is the same as that of Figure~\ref{fig:rho}.
Radial velocity vectors are overplotted to show the direction of flow relative to the clump.}
\label{fig:clumpsrho}
\end{figure}

We now turn to the behavior of the working surfaces which are the main focus of this study.
The effect of the magnetic field hoop stresses is readily seen in the development of high density axial knots discussed above.
These are expected based on the analysis in Section~\ref{sec:theory}.
In Figure~\ref{fig:clumpsrho} we present an enlarged view of a clump seen at the same time for each model.
Note how the clump width $r_{cl}$ decreases with increasing field strength (decreasing $\beta$).
In addition, the ``height'' of the clump ($z_{cl}$), measured as the distance between the forward and reverse shocks of the internal working surfaces along the axis, increases with increasing field strength (decreasing $\beta$).
The decrease in $r_{cl}$ and the increase in $z_{cl}$ can be attributed to the effect of the magnetic hoop stress $F_h$ and magnetic pressure $F_{bp}$ respectively.
Hoop stresses act to restrict lateral flow leading to a build-up of torodial field on the axis.
The increase in $B_\phi$ then produces a pressure gradient $\nabla B_\phi$ which forces the clump to expand both upstream and downstream in the z-direction.

The effect of $F_h = (\frac{B^2}{4 \pi R})\hat{\bf e}_r$ can be seen in Figure~\ref{fig:clumpsrho} which shows the direction of the radial velocity vectors $v_r$ within the working surface (i.e. compression or expansion).
In these images we show only the direction of the flow vectors and filter out gas parcels with $v_r < 0.1 v_{r,max}$.
This allows us to see the strong radial flows driven by magnetic and gas pressure forces.
The direction of the radial velocity vectors confirms that material is compressed towards the jet axis for the MHD cases while the $\beta = \infty$ case shows little lateral motion in the axial region.
Figure~\ref{fig:clumpsrho} also confirms the analysis of Section~\ref{sec:theory} in that we see the radial velocity change from axial expansion to compression within the clump as a function of radius.
Based on the analysis from Section~\ref{sec:theory}, we would expect the transition to occur at $R_m$, or 0.6$R_j$.
This is consistent with what is seen in our simulations though the bifurcation occurs in the simulations at smaller radii, $r \sim 0.4 R_j$.
Note that our theoretical analysis relied on the assumption that the pre-shock conditions are the jet injection parameters.
As shown by \citet{Gardiner00}, however, the torodial field in the undisturbed jet beam will strengthen before a shock forms as each pulse steepens.
The field evolution follows

\begin{equation}
B_{\phi} = \frac{B_{\phi,0}}{[1-\kappa(1-t_0)]} ,\label{11}
\end{equation}
where $\kappa$ is the ratio of the derivative of the jet velocity to the jet velocity at the time $t_0$ when the gas parcel under consideration was launched $\kappa=v_j'(t_0)/v_j(t_0)$.
Thus the field will already be strengthened as the pulses steepen meaning that a higher field value should be used in the shock analysis.
This is consistent with our observation that the bifurcation radius where behavior switches from expansion to compression occurs at a slightly smaller radius.

\begin{figure}
\epsscale{0.8}
\plotone{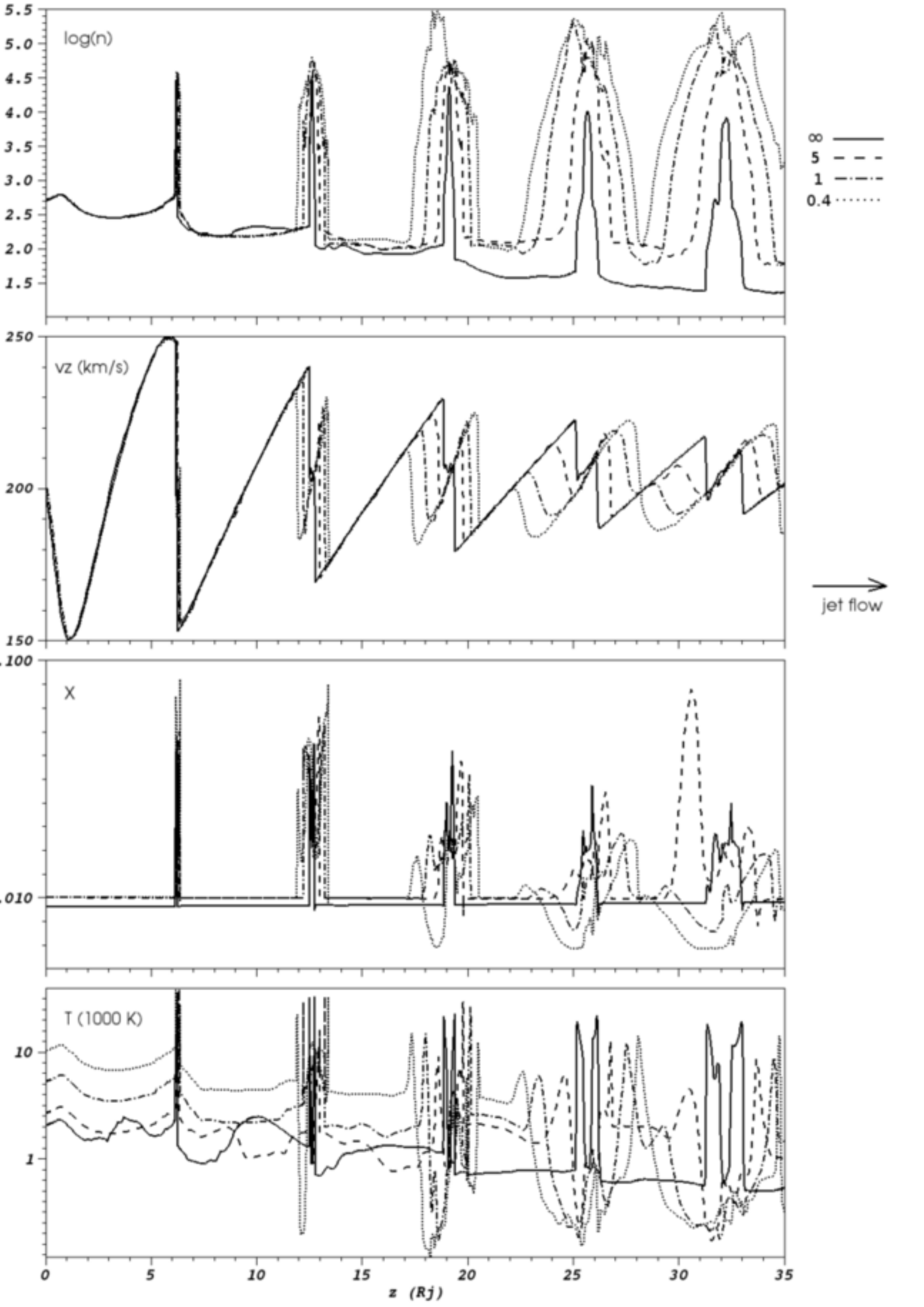}
\caption{Working surfaces within the pulsed, magnetized jets.
The axial lineouts show the logarithm of number density (cm\textsuperscript{-3}), axial velocity, ionization fraction, and temperature from \emph{top} to \emph{bottom} respectively.
The $\beta = \infty, 5, 1, 0.4$ models are represented by the \emph{solid}, \emph{dashed}, \emph{dot-dahsed}, and \emph{dotted} lines respectively.
The lineouts are taken at $r = 0.1 R_m$ and $t = 500$ yrs.}
\label{fig:lineouts}
\end{figure}

We now consider profiles of density, axial velocity, ionization fraction, and temperature taken along the jet axis (at $r = 0.1 R_m$) after the jet has propagated for 500 years.
In these plots, shown in Figure~\ref{fig:lineouts}, the \emph{solid}, \emph{dashed}, \emph{dash-dotted}, and \emph{dotted} lines represent the $\beta = \infty, 5, 1, 0.4$ runs respectively.
The jet is propagating to the right in this figure and the head of the jet is not shown.
The profiles of density along the jet axis (Figure~\ref{fig:lineouts}) follow the expected pattern in which the region between the working surface shock-pairs becomes denser and broader ($z_{cl}$ increases) as the pulse propagates downstream.
In addition to this generic behavior, we see the $\rho_{cl}$ and $z_{cl}$ increase with increasing field strength.
The increased density occurs to the increased compression driven by $B_\phi$ tension (pinch) forces while the increase in $z_{cl}$ occurs due to $B_\phi$ pressure forces

The $v_z$ profiles show similar behavior as a function of $B_{\phi,0}$.
Note that the pulse closest to the jet source still shows its initial sine wave shape truncated by a shock-pair which has formed at $z \sim 7R_j$.
As the pulses propagate downstream, the distance between the upstream and downstream facing shocks increases.
The internal shocks also weaken over time which increases the cooling length.
At later times, the discontinuities first smooth and then re-steepen as the reservoir of high velocity material is used up until mergers with other clumps lead to new shocks.

The temperature profiles show the strong shock jump at the edges of the clumps.
Note that our equilibrium conditions in the initial jet imply higher values of temperature in the undisturbed jet beam.
Given the amplitude of the velocity pulses, we expect shocks of strength $v_s \sim 50$ km/s \citep{Hartigan01} which should produce post-shock temperatures of $T_s = \frac{3 \mu m_H}{16 k_b} v_s^2 \sim 10^4 K$.
The maximum temperature seen behind our shocks is $T_s \sim 51000$ K which demonstrates that our resolution is high enough to capture the full run of post-shock heating and cooling.
Within each clump we can, therefore, see the presence of a cooling zone driven by ionized species such as SII.
It is noteworthy that the simulations with higher magnetic fields tend to produce lower temperatures within the clump.
This is essentially an effect of the hoop stresses producing higher densities in the clump via compression.
Since cooling can be expressed as $L \sim n^2 \Lambda(T)$, where $\Lambda(T)$ is the cooling curve for each line, higher number densities will lead to stronger cooling.
In the working surfaces, densities will continue to increase until radial pressure forces become strong enough to counteract the hoop stresses.

The ionization fraction profiles follow temperature and density profiles behind the shocks as expected given the functional form of the ionization rate equations are 

\begin{equation}
\frac{d n_{H_0}}{dt} = \zeta_{H_+}(T)n_en_{H_+} - \Gamma_{H_0}(T)n_en_{H_0} ,\label{12}
\end{equation}
where $\zeta$ and $\Gamma$ are the recombination and collisional ionization conditions respectively.
While we begin with $x = n_{H_+}/n_H = 0.01$ we see the increase in ionization behind shocks and the subsequent decrease in the cooling zones.
Once again the axial compression driven by strong fields produces higher values of $n_e$ and $n_{H_+}$ leading to strong recombination and the lowest values of $x$ in all simulations ($x \sim 0.006$).

\begin{figure}
\plotone{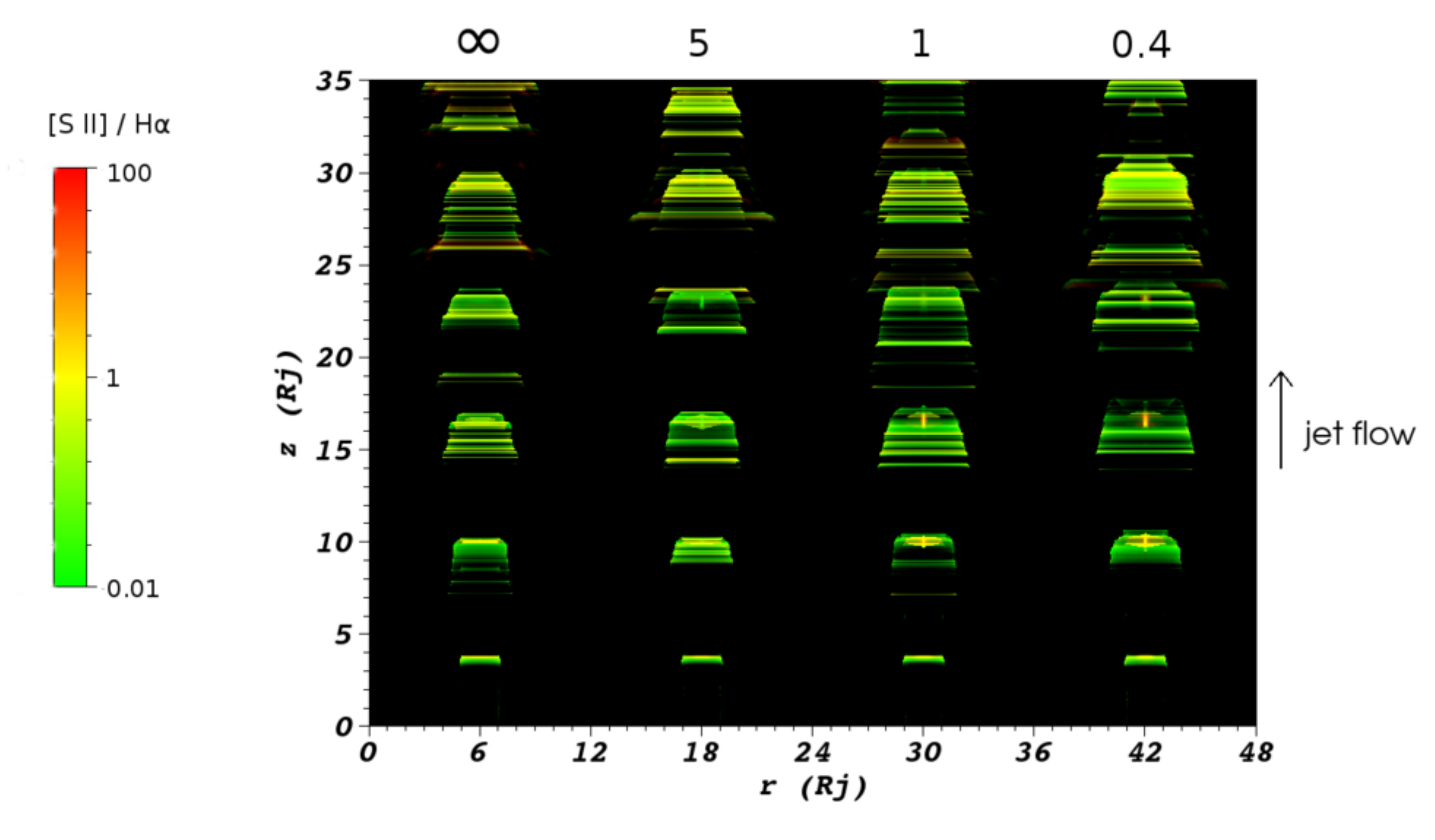}
\caption{Effect of magnetic field on emission line structure of the internal clumps in pulsed jets.
These emission maps show H$\alpha$ and [S II] for the $\beta = \infty, 5, 1, 0.4$ models (from \emph{left} to \emph{right} respectively).
Green corresponds to H$\alpha$ emission and red is the [S II].
The head of the jet has been excluded for simplicity (axial domain from $z = 0$ to $z = 35 R_j$)
Both the H$\alpha$ and [S II] emission are scaled by a factor of $10^{-18}$ at this simulation time ($t = 480$ yr for these images).
The RGB scale (from 0 to 255) spans two orders of magnitude for each emission line logarithmically, which results in the shown scale.} 
\label{fig:emiss}
\end{figure}

We now turn to the observational consequences of the dynamics seen in our simulations.
In Figure~\ref{fig:emiss} we present projected images in H$\alpha$ and [S II] where green corresponds to H$\alpha$ emission and red is the [S II].
Yellow marks regions that have both strong H$\alpha$ and [S II] emission.
These images are computed by rotating the 2.5-D emission maps and computing line of site integrations through the resulting 3-D emission distributions.
The data are all shown with the jet axis perpendicular to the plane of the sky.
The most evident aspect of the images is the cascade of shocks associated with each clump.
Moving from the jet inlet to the jet head, we see an increasing complex set of nested ``bow'' or ``wing'' shaped shock features.
We first note that it is the rotation of the axisymmetric emission maps the smooth and highly symmetric nature of the observed patterns.
In spite of the imposed axisymmetry, we can attribute the complexity of the observed emission pattern with increasing height to the turbulent flows associated with each working surface captured by our higher resolution runs.
Compare these images to, for example, those in \citet{deColle06} in which a resolution of 0.22 cells/$L_{cool}$ was used, and the resulting images show fewer bow/wing features.

Comparison of the 4 simulations does not yield any immediate distinction in terms of the effect of the field on the pattern of bow shock emission; there are multiple nested features in all cases.
In all cases we see bright H$\alpha$ appearing at the head of the strongest shocks (as expected) with [S II] emission appearing behind the shocks in regions that are dominated by cooling.
What is, however, quite apparent is the presence of bright axial features in H$\alpha$ in the runs with stronger fields.
In the $\beta = \infty$ run these features are non-existent while in the $\beta = 0.4$ run they traverse the extent (in $z$) of each clump.

\begin{figure}
\plotone{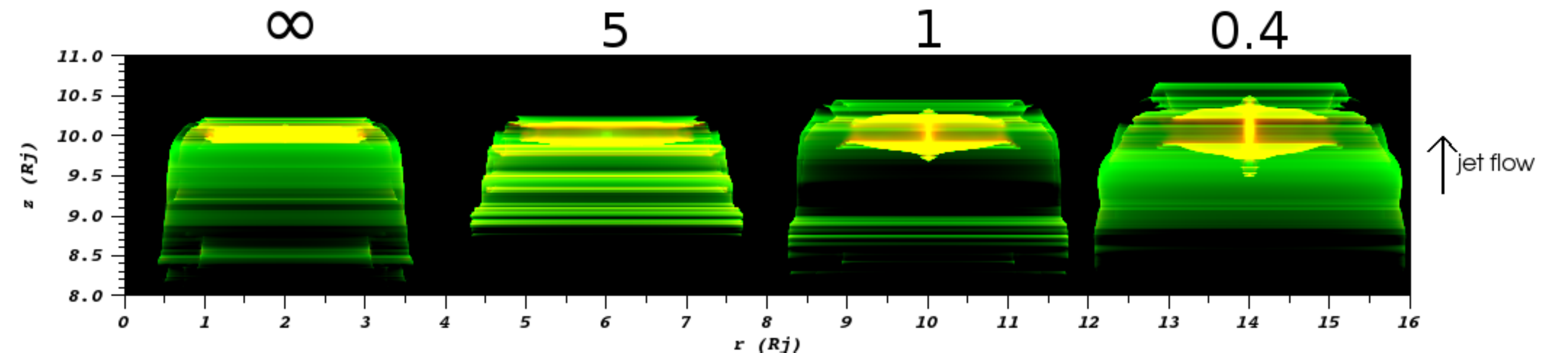}
\caption{Effect of magnetic field on emission line structure of a clump.
These are enlarged H$\alpha$ and [S II] emission maps for all 4 models showing the same clump located near $z = 10 R_j$ at $t = 480$ yrs.
Scaling is the same as in Figure~\ref{fig:emiss}.}
\label{fig:clumpemiss}
\end{figure}

To explore these features in greater detail we present in Figure~\ref{fig:clumpemiss} synthetic images of the same clumps shown in Figure~\ref{fig:clumpsrho}.
We see the magnetic field affecting emission in two ways.
First it alters post-shock conditions; since the magnetic field represents another source of pressure we see an extension of the cooling zones.
In this way, the field acts as a ``shock absorber'' resulting in broader [S II] emission.
Hence, the red regions of [S II] are extended farther in the MHD cases when compared to the $\beta = \infty$ case in Figure~\ref{fig:clumpemiss}.
The second effect comes from the hoop stresses driving material towards the axis.
As we move from the $\beta = \infty$ simulation to the strongest field $\beta = 0.4$ run we see the axial compression producing a bright cylindrical region of H$\alpha$ emission.
Thus the axial flow driven by the hoop stresses produces a cylindrical shock embedded within the working surface.
The larger the initial field in the simulation, the brighter and longer the resulting H$\alpha$ ``core''.
Note that in the $\beta = 1$ and $\beta = 0.4$ cases, the core has punched through the working surface in both the up and downstream directions creating its own small scale version of a jet with bow-shocks in both directions.
The existence of the H$\alpha$ core is the most prominent new feature seen in our simulations.
It is not clear if such a feature could occur in a 3-D simulation.
Such toroidal-field dominated features are likely to be destabilized by m=1 kink modes which would turn the core into a sequence of knots.
The knots however may end up with an observable signature due to the presence of strong fields.

\begin{figure}
\plotone{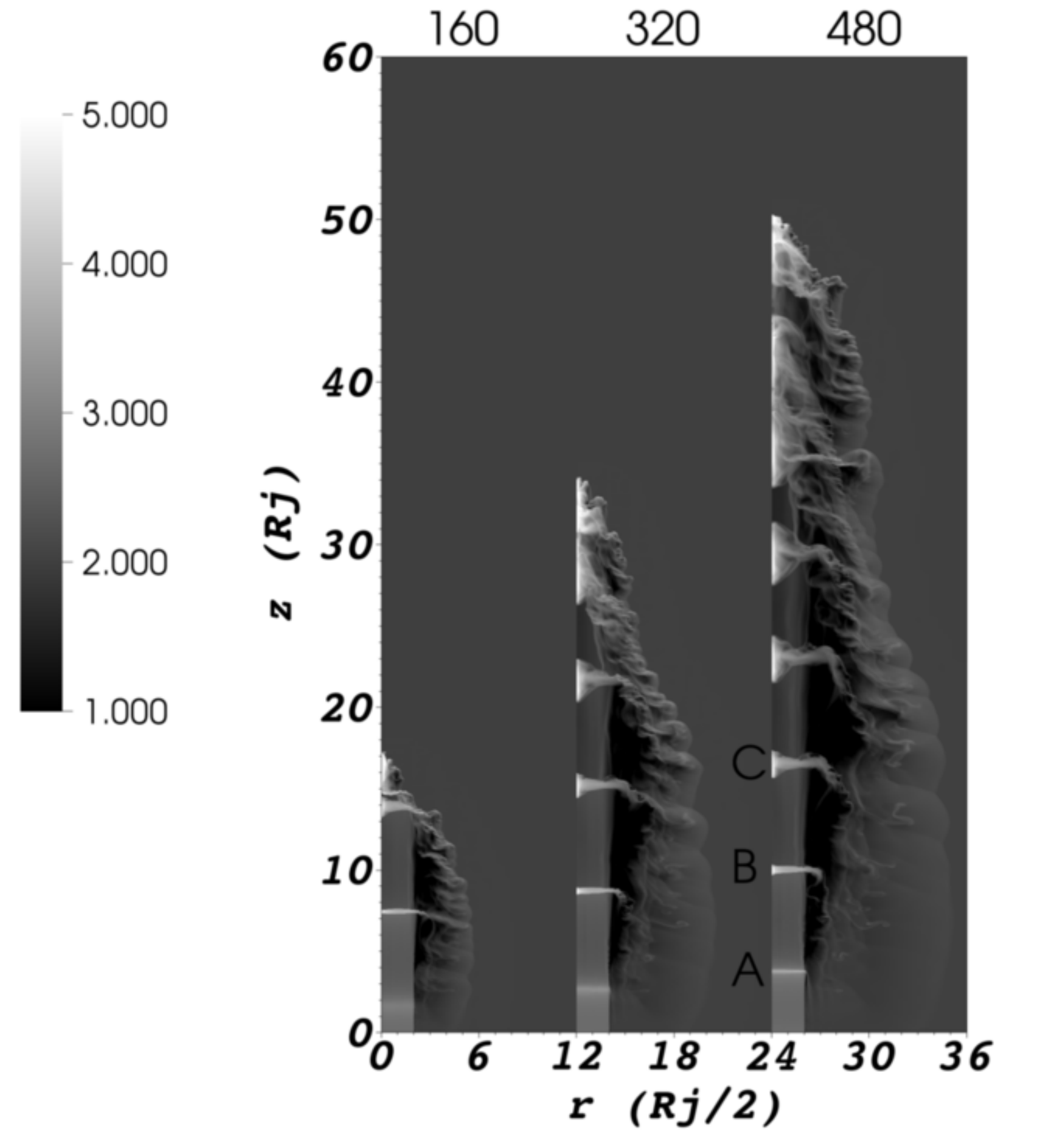}
\caption{Time evolution of axial density of a pulsed, magnetized jet.
This shows a time series of the number density (cm\textsuperscript{-3}) of the $\beta = 1$ model.
The panels, from \emph{left} to \emph{right}, correspond to times of 160, 320, and 480 years respectively.}
\label{fig:beta1jet}
\end{figure}

To clarify the time evolution of the flow in Figure~\ref{fig:beta1jet} we present the  $\beta = 1$ simulation at 3 different times.
This figure demonstrates two important points.
First we see how flows from the edges of the working surface feed into the jet cocoon.
As noted earlier, the high resolution in our simulation allows us to capture the fragmentation of these flows which are likely due to a variety of instabilities.
We see that as each pulse propagates up the jet beam, shocked material near $R_j$ is driven (by thermal pressure) laterally away from the jet where it encounters a mix of ambient gas and material ejected from previous working surfaces.
New shocks form (which represent the bows seen in Figure~\ref{fig:emiss}), and the flow becomes turbulent.
Note that our refinement criterion for the simulation are such that the flows become de-refined at larger distances (and weaker shocks) from the jet boundary.
The second point to note is in relation to the evolution of the clumps near the axis.
Figure~\ref{fig:beta1jet} lets us see the formation of high densities due to the action of the hoop stresses.
The youngest clump (A) in the image remains thin and planar.
By the time a pulse reaches the position of clump B, we can already see the beginnings of the axial condensation that has begun to expand upstream and downstream.
Clump C shows the continual expansion (in $z$) of the axial core.

\begin{figure}
\plotone{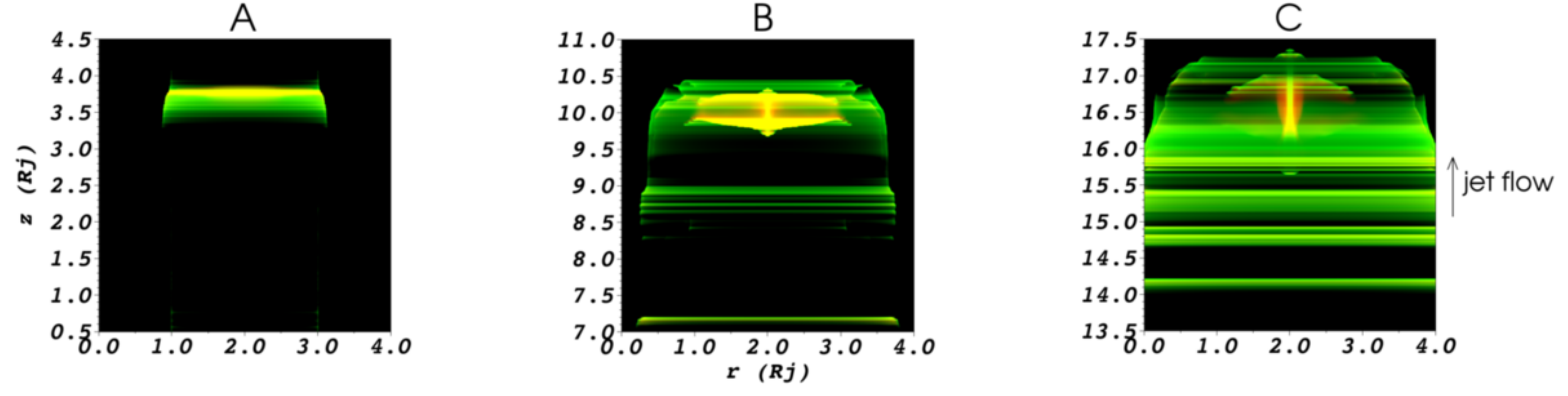}
\caption{Time evolution of clumps in a pulsed, magnetized jet.
These are enlarged images of the clumps denoted in Figure~\ref{fig:beta1jet}.
This figure shows the emission maps of H$\alpha$ (green) and [S II] (red).
The scaling is the same as in Figures~\ref{fig:emiss}~and~\ref{fig:clumpemiss}.
The ages of clumps A, B, and C are approximately 30, 80, and 130 yrs respectively.}
\label{fig:beta1clumpemiss}
\end{figure}

Figure~\ref{fig:beta1clumpemiss} presents synthetic observations of these three clumps.
Clump A shows upper and lower shocks both in bright H$\alpha$ features with some [S II] region in between.
Clump B shows broadening of the [S II] cooling region and the signature appearance of the bright H$\alpha$ core. 
In the more evolved clump C, we see the [S II] emission fading as the interior material cools.
The H$\alpha$ continues to expand in $z$.
Notably clump C shows diffuse H$\alpha$ emission.
This marks a shock at the edge of the entire working surface ($r > R_j$) formed as thermal pressure drove material into the cocoon.

Thus our simulations show a well ordered patterned of evolution within strongly magnetized axisymmetric working surfaces in terms of both the flow pattern and their observable consequences.
The dominant effect of the fields is the formation of the axial cores which form mini-jets that expand into the upstream and downstream flow.
Magnetic pressure across the entire working surface, however, still acts to broaden them and expand the cooling region dominated by [S II] emission.

\section{A Randomized Velocity Model}
\label{sec:random}
The sinusoidally varying injection velocity used in the previous simulations creates evenly spaced knots within the jet beam.
However observations often show knots that are not evenly spaced with more knots closer to the source \citep{Hartigan11}.
One way of achieving such a distribution of knots is via a randomized injection velocity \citep{Raga92}.
We have therefore also completed a simulation with such random pulse behavior to explore its differences from the sinusoidally pulsed model.

The method for generating a randomized injection velocity is the same as used by \citet{Hartigan07}.
The input velocity is a series of random steps such that

\begin{equation}
v(t) = v_0 (1 + A k) ,\label{13}
\end{equation}
where $v_0$ is 200 km/s and $A = 0.25$ as before.
The variable $k$ is a random number between -1 and 1, and it changes every 5 years within the simulation.
For this simulation, the domain size has been increased in the $z$ direction, but the resolution is the same as before.
The run time has also been increased from 500 years to 600 years.
We only ran one model with $\beta = 1$, as we did not wish to conduct another parameter study.
All other parameters are the same as the sinusoidal velocity simulations.

\begin{figure}
\plotone{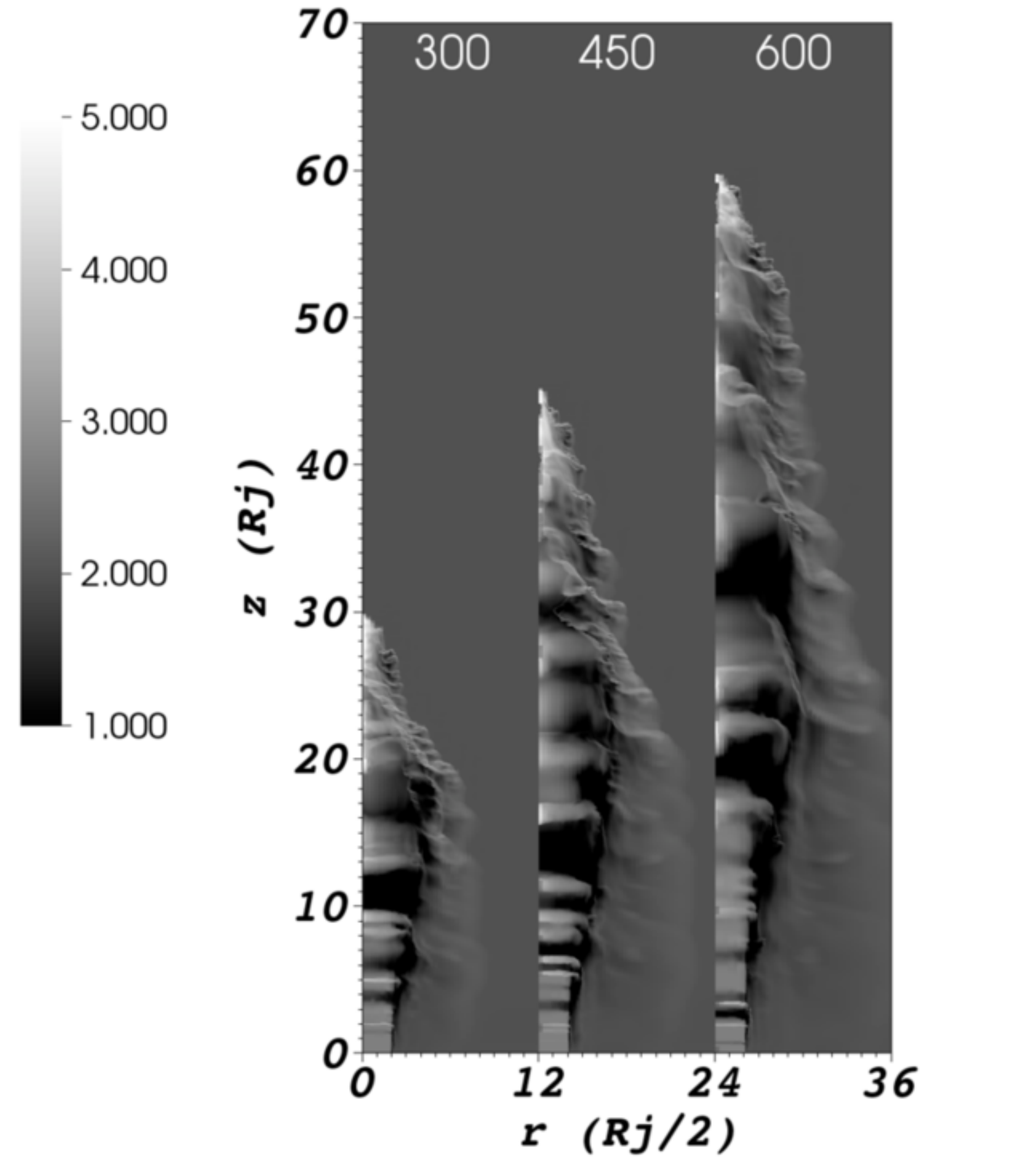}
\caption{Time evolution of axial density of a magnetized jet with random velocity pulses.
This shows a time series of the number density (cm\textsuperscript{-3}) of a random injection velocity model with $\beta = 1$.
The panels, from \emph{left} to \emph{right}, correspond to times of 300, 450, and 600 years respectively.}
\label{fig:randrho}
\end{figure}

Figure~\ref{fig:randrho} presents a time series of the axial density showing the overall morphology of the jet.
It is immediately obvious that the structure has become more complex.
There are indeed more clumps near the jet source than there are farther away, and the clumps that are farther away are not evenly spaced.
As in the other models, the fast moving clumps catch up to the head of the jet and terminate there.
These clumps contribute to the jet head density and momentum, pushing it farther and farther into the ambient.
We again see the formation of a nose-cone in this $\beta = 1$ model, and it is especially prominent at 450 years.
The fact that the nose-cone looks steeper at 450 years than at 600 years suggests that a mechanism within the jet head can disrupt the nose-cone.
This can be due to the cooling instabilities present there as well as the colliding working surfaces.

The knots in Figure~\ref{fig:rho} appear denser than those in Figure~\ref{fig:randrho}.
When the pulses are random, the region close to the source contains many shocks and rarefactions of varying strengths.
These shocks and rarefactions interact with each other, and some of the interactions form dense working surfaces that are observed farther downstream.
However most of the ineractions lead to moderate densities within weaker shocked or rarefaction regions that get ``smeared'' out as they propagate through the jet material.
In a sinusoidally pulsed flow, the injected material has more time to combine and form a dense working surface, with stronger shocks, that propagates through the jet beam.

\begin{figure}
\epsscale{0.8}
\plotone{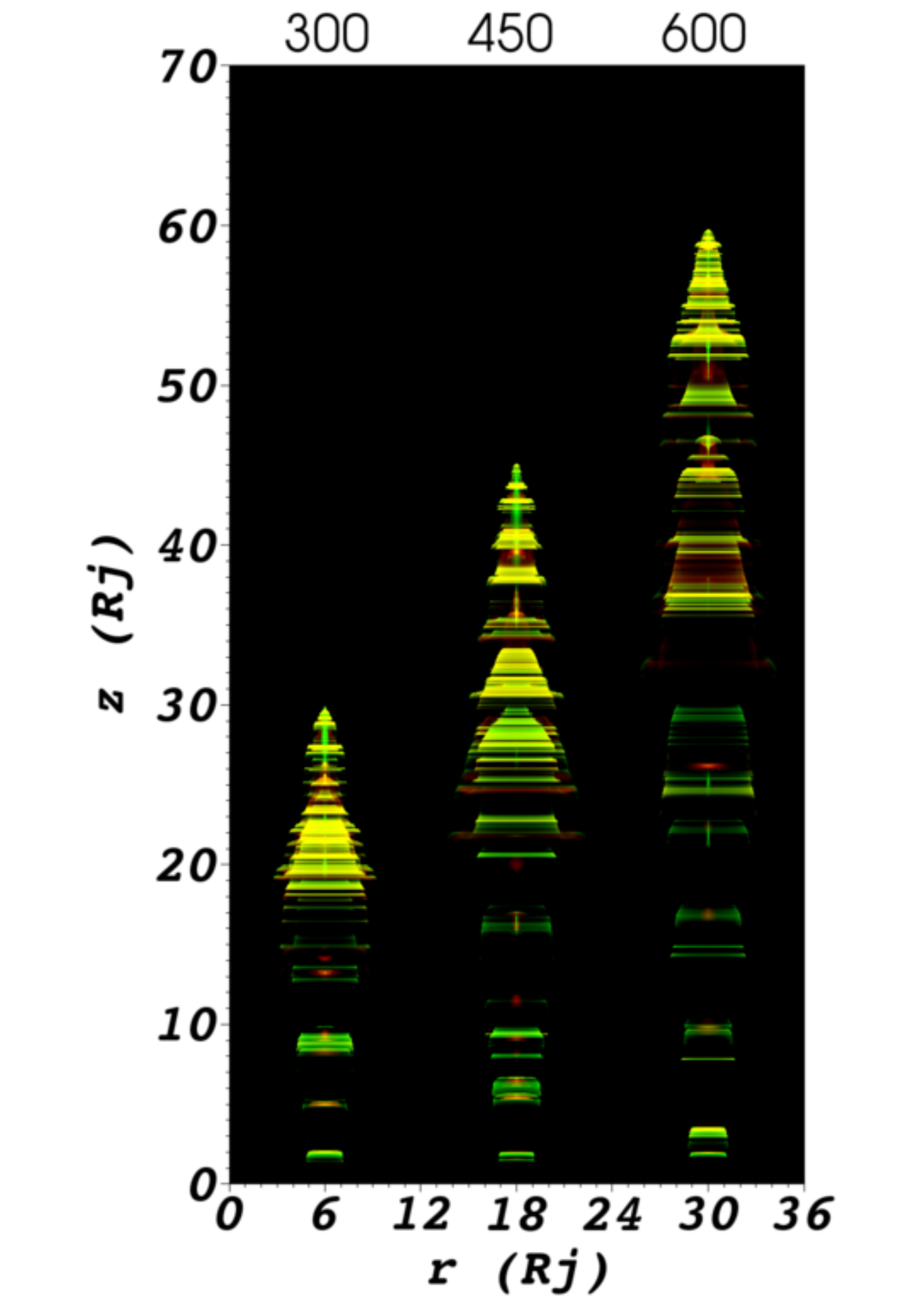}
\caption{Time evolution of emission line structure of a magnetized jet with random velocity pulses.
This shows a time series of the H$\alpha$ (green) and [S II] (red) emission of a random injection velocity model with $\beta = 1$.
The panels, from \emph{left} to \emph{right}, correspond to times of 300, 450, and 600 years respectively.
The scaling is the same as in the previous emission map figures.}
\label{fig:randemiss}
\end{figure}

Figure~\ref{fig:randemiss} shows the same time series for the H$\alpha$ and [S II] emission.
The picture here is consistent with the aformentioned interpretation.
We see the time-dependnet nose-cone form at the head of the jet, and we see an uneven spacing of knots.
Due to the randomness of the injected velocities, the region near the source does not always appear to have more knots.
This is because the strengths of the shocks here are different at different times.
Nevertheless, Figure~\ref{fig:randrho} clearly shows that there are indeed more dense clumps closer to the source.

As in the previous models, we see that the emission has a complicated time dependence.
There are H$\alpha$ and [S II] regions that brighten and fade over time which is consistent with observations of HH objects \citep{Hartigan11}.
We also again see H$\alpha$ features widening as an internal working surface propagates downstream.
As with the sinusoidally driven jets this is due to secondary bow shocks, or ``spur'' shocks, that move radially outwards despite the strong magnetic hoop stresses present.

\section{Discussion and Conclusions}
\label{sec:discconc}
We have completed high resolution 2.5-D MHD simulations of pulsed YSO jets.
Four cases were studied: $\beta = \infty$ (no field), and $\beta = 5, 1, 0.4$.
Increasing the magnetic field strength demonstrated a progression of behavior in which field tension and field pressure play an increasingly important role in the jet evolution.

Resolving the cooling length is crucial to capturing the radiative processes and thus the emission (\citet{Yirak10}).
The resolution of our simulations represent some of the highest yet applied to MHD jets studies (\citet{Tesileanu12,deColle06}).
By using a non-equilibrium cooling curve based on detailed shock models we have also been able to calculate both H$\alpha$ and [S II] synthetic emission maps.
We also checked for numerical convergence, and a resolution study confirmed that the basic morphological features of the simulations had converged.

Our theoretical analysis, building off that of \citet{deColle08}, demonstrated the mach number and radial position dependence of pinch forces in radiative MHD jets.
The simulations confirmed our predictions for the variation of tension forces within internal working surfaces.
Stronger magnetic fields lead to both internal clump compression in the radial direction via hoop stresses and axial clump broadening due to magnetic pressure forces.
Thus the time dependence of magnetic pinch force leads to the formation of a compressed core in the working surfaces which gradually increases in extent in the direction parallel to the jet propagation. 
The simulations also demonstrate the role of such pinch forces at the jet head.
We find an increase in magnetic field strength is consistent with an increase in axial jet head velocity in all cases.

The theoretical framework presented in Section~\ref{sec:theory}, confirmed in the simulations, also accounts for the variation of emission patterns within the working surfaces.
In particular the pinch forces lead to bright H$\alpha$ emission surrounding the knot cores embedded in [S II] regions between the internal shocks.

Our random pulse simulation showed similar behavior within the working surfaces as we saw in the periodic simulations.
As expected, however, the eventual merger of faster shocks downstream leads to a global morphology for the random pulse model with more working surfaces closer to the jet source.

We note that our work was limited to a study of toroidal magnetic fields.
In future work it may be interesting to study other field geometries such as helical fields which have been shown to be important as they can drive jet rotation (\citet{Fendt11}).

Further work is also needed to resolve the issue of nose-cones.
The first discussions of these features was presented almost 30 years ago \citet{Clarke86} and still the true nature of such features in real jets remains unclear.
It is expected that nose-cones result from the imposed symmetry in 2.5-D simulations.
While the analysis of Section~\ref{sec:theory} implies that the pinch forces producing ``cores'' and the nose-cone should be present in 3-D, it is likely that kink instabilities will disrupt the formation of such toroidal-field dominated structures.  

The use of 2.5-D simulations clearly limits the connections we can draw with observations.
The simulations, however, do produce the patterns of H$\alpha$ and [S II] expected from shock pairs in flows with velocity variations and they also demonstrate the importance of magnetically modified ``bow'', ``wing'', or ``spur'' shock patterns which form as material is ejected transverse to the flow.
These conclusions are unlikely to change significantly in 3-D.

With regard to observations, perhaps the most relevant result pertains to the dominance of pinch forces in creating the bright cores.
This implies a mechanism by which clumps can naturally form in the jets.
One of the principle conclusions of work done by \citet{Hartigan11} was the existence of multiple sub-radial sized structures in the jet beam (i.e. clumps).
The toroidally dominated cores in our simulations will be subject to kink mode instabilities which laboratory experiments have also shown resolve into multiple clumps with a range of velocities relative to the bulk flow \citep{Ciardi07}.
These groups of clumps will propagated downstream and, perhaps, lead to the dramatic ``shock curtains'' apparent in the multi-epoch observations of YSO jets.

Thus our work sets the stage for 3-D studies of radiative MHD jets.
Since those global studies will have to run at lower resolution the current work provides a baseline for fully resolved multi-dimensional cooling flows in jets.

\vspace{7 mm}
\noindent\emph{Acknowledgements.} This work used the Extreme Science and Engineering Discovery Environment (XSEDE), which is supported by National Science Foundation grant number OCI-1053575. The CIRC at the University of Rochester provided computational resources. Financial support for this project was provided by the LLE at the University of Rochester, Space Telescope Science Institute grants HST-AR-11251.01-A and HST-AR-12128.01-A; by the National Science Foundation under award AST-0807363; by the Department of Energy under award de-sc0001063.  A special thanks to Baowei Liu, Jonathan Carroll-Nellenback, and Martin Huarte-Espinosa for their assistance with using the AstroBEAR code and running the simulations.

\bibliographystyle{apj}
\bibliography{mylibrary}

\begin{thebibliography}{43}
\expandafter\ifx\csname natexlab\endcsname\relax\def\natexlab#1{#1}\fi

\bibitem[{Blondin \& Marks(1996)}]{Blondin96}
Blondin, J.~M. \& Marks, B.~S. 1996, NewA, 1, 235

\bibitem[{Carroll-Nellenback {et~al.}(2012)Carroll-Nellenback, Shroyer, Frank,
  \& Ding}]{Carroll12}
Carroll-Nellenback, J., Shroyer, B., Frank, A., \& Ding, C. 2012, ASPC, 459,
  291

\bibitem[{Cerqueira \& {de Gouveia Dal Pino}(2001)}]{Cerqueira01}
Cerqueira, A.~H. \& {de Gouveia Dal Pino}, E.~M. 2001, ApJ, 560, 779

\bibitem[{Ciardi {et~al.}(2007)Ciardi, Lebedev, Frank, Blackman, Chitendden,
  Jennings, Ampleford, Bland, Bott, Rapley, Hall, Suzuki-Vidal, Marocchino,
  Lery, \& Stehle}]{Ciardi07}
Ciardi, A., Lebedev, S.~V., Frank, A., Blackman, E.~G., Chitendden, J.~P.,
  Jennings, C.~J., Ampleford, D.~J., Bland, S.~N., Bott, S.~C., Rapley, J.,
  Hall, G.~N., Suzuki-Vidal, F.~A., Marocchino, A., Lery, T., \& Stehle, C.
  2007, Physics of Plasmas, 14, 056501

\bibitem[{Clarke {et~al.}(1986)Clarke, Norman, \& Burns}]{Clarke86}
Clarke, D.~A., Norman, M.~L., \& Burns, J.~O. 1986, ApJ, 311, L63

\bibitem[{Cunningham {et~al.}(2009)Cunningham, Frank, Varni{\`e}re, Mitran, \&
  Jones}]{Cunningham09}
Cunningham, A.~J., Frank, A., Varni{\`e}re, P., Mitran, S., \& Jones, T.~W.
  2009, ApJS, 182, 519

\bibitem[{Dalgarno \& McCray(1972)}]{Dalgarno72}
Dalgarno, A. \& McCray, R.~A. 1972, A\&AA, 10, 375

\bibitem[{{de Colle} \& Raga(2006)}]{deColle06}
{de Colle}, F. \& Raga, A.~C. 2006, A\&A, 449, 1061

\bibitem[{{de Colle} {et~al.}(2008){de Colle}, Raga, \& Esquivel}]{deColle08}
{de Colle}, F., Raga, A.~C., \& Esquivel, A. 2008, ApJ, 689, 302

\bibitem[{Fendt(2011)}]{Fendt11}
Fendt, C. 2011, ApJ, 737, 43

\bibitem[{Frank {et~al.}(1998)Frank, Ryu, Jones, \& Noriega-Crespo}]{Frank98}
Frank, A., Ryu, D., Jones, T.~W., \& Noriega-Crespo, A. 1998, ApJ, 494, L79

\bibitem[{Gardiner \& Frank(2000)}]{Gardiner00}
Gardiner, T.~A. \& Frank, A. 2000, ApJ, 545L, 153G

\bibitem[{Hartigan {et~al.}(2011)Hartigan, Frank, Foster, Wilde, Douglas,
  Rosen, Coker, Blue, \& Hansen}]{Hartigan11}
Hartigan, P., Frank, A., Foster, J.~M., Wilde, B.~H., Douglas, M., Rosen,
  P.~A., Coker, R.~F., Blue, B.~E., \& Hansen, J.~F. 2011, ApJ, 736, 29

\bibitem[{Hartigan {et~al.}(2007)Hartigan, Frank, Varni{\'e}re, \&
  Blackman}]{Hartigan07}
Hartigan, P., Frank, A., Varni{\'e}re, P., \& Blackman, E.~G. 2007, ApJ, 661,
  910

\bibitem[{Hartigan {et~al.}(2014)Hartigan, Hansen, Wright, \&
  Frank}]{Hartigan15}
Hartigan, P., Hansen, E.~C., Wright, A., \& Frank, A. 2014, in preparation

\bibitem[{Hartigan {et~al.}(2005)Hartigan, Heathcote, Morse, Reipurth, \&
  Bally}]{Hartigan05}
Hartigan, P., Heathcote, S., Morse, J.~A., Reipurth, B., \& Bally, J. 2005, AJ,
  130, 2197

\bibitem[{Hartigan {et~al.}(1994)Hartigan, Morse, \& Raymond}]{Hartigan94}
Hartigan, P., Morse, J.~A., \& Raymond, J. 1994, ApJ, 436, 125

\bibitem[{Hartigan {et~al.}(2001)Hartigan, Morse, Reipurth, Heathcote, \&
  Bally}]{Hartigan01}
Hartigan, P., Morse, J.~A., Reipurth, B., Heathcote, S., \& Bally, J. 2001,
  ApJ, 559, L157

\bibitem[{Hartigan \& Raymond(1993)}]{Hartigan93}
Hartigan, P. \& Raymond, J. 1993, ApJ, 409, 705

\bibitem[{Hartigan {et~al.}(1987)Hartigan, Raymond, \& Hartmann}]{Hartigan87}
Hartigan, P., Raymond, J., \& Hartmann, L. 1987, ApJ, 316, 323

\bibitem[{Heathcote {et~al.}(1996)Heathcote, Morse, Hartigan, Reipurth,
  Schwartz, Bally, \& Stone}]{Heathcote96}
Heathcote, S., Morse, J.~A., Hartigan, P., Reipurth, B., Schwartz, R.~D.,
  Bally, J., \& Stone, J.~M. 1996, AJ, 112, 1141

\bibitem[{Kajdi{\u c} {et~al.}(2006)Kajdi{\u c}, Vel{\'a}zquez, \&
  Raga}]{Kajdic06}
Kajdi{\u c}, P., Vel{\'a}zquez, P.~F., \& Raga, A.~C. 2006, RevMexAA, 42, 217

\bibitem[{K{\"o}ssl {et~al.}(1990{\natexlab{a}})K{\"o}ssl, M{\"u}ller, \&
  Hillebrandt}]{Kossl90a}
K{\"o}ssl, D., M{\"u}ller, E., \& Hillebrandt, W. 1990{\natexlab{a}}, A\&A,
  229, 378

\bibitem[{K{\"o}ssl {et~al.}(1990{\natexlab{b}})K{\"o}ssl, M{\"u}ller, \&
  Hillebrandt}]{Kossl90b}
---. 1990{\natexlab{b}}, A\&A, 229, 397

\bibitem[{Lind {et~al.}(1989)Lind, Payne, Meier, \& Blandford}]{Lind89}
Lind, H., Payne, D., Meier, D., \& Blandford, R. 1989, ApJ, 344, 89

\bibitem[{Mignone {et~al.}(2010)Mignone, Rossi, Bodo, Ferrari, \&
  Massaglia}]{Mignone10}
Mignone, A., Rossi, P., Bodo, G., Ferrari, A., \& Massaglia, S. 2010, MNRAS,
  402, 7

\bibitem[{Norman {et~al.}(1982)Norman, Smarr, Winkler, \& Smith}]{Norman82}
Norman, M.~L., Smarr, L., Winkler, K.-H.~A., \& Smith, M.~D. 1982, A\&A, 113,
  285

\bibitem[{O'Sullivan \& Ray(2000)}]{OSullivan00}
O'Sullivan, S. \& Ray, T.~P. 2000, A\&A, 363, 355

\bibitem[{Raga(1992)}]{Raga92}
Raga, A.~C. 1992, MNRAS, 258, 301

\bibitem[{Raga {et~al.}(1990)Raga, Binette, Canto, \& Calvet}]{Raga90}
Raga, A.~C., Binette, L., Canto, J., \& Calvet, N. 1990, ApJ, 364, 601

\bibitem[{Raga {et~al.}(1991)Raga, Biro, Cant{\'o}, \& Binette}]{Raga91}
Raga, A.~C., Biro, S., Cant{\'o}, J., \& Binette, L. 1991, RevMexAA, 22, 243

\bibitem[{Raga {et~al.}(2007)Raga, {de Colle}, Kajdi{\u c}, Esquivel, \&
  Cant{\'o}}]{Raga07}
Raga, A.~C., {de Colle}, F., Kajdi{\u c}, P., Esquivel, A., \& Cant{\'o}, J.
  2007, A\&A, 465, 879

\bibitem[{Raga {et~al.}(2002{\natexlab{a}})Raga, Noriega-Crespo, Reipurth,
  Garnavich, Heathcote, B{\"o}hm, \& Curiel}]{Raga02ApJ}
Raga, A.~C., Noriega-Crespo, A., Reipurth, B., Garnavich, P.~M., Heathcote, S.,
  B{\"o}hm, K.~H., \& Curiel, S. 2002{\natexlab{a}}, ApJ, 565L, 29

\bibitem[{Raga {et~al.}(2012)Raga, Rodriguez-Gonz{\'a}lez, Noriega-Crespo, \&
  Esquivel}]{Raga12}
Raga, A.~C., Rodriguez-Gonz{\'a}lez, A., Noriega-Crespo, A., \& Esquivel, A.
  2012, ApJ, 744L, 12

\bibitem[{Raga {et~al.}(2002{\natexlab{b}})Raga, Vel{\'a}zquez, Cant{\'o}, \&
  Masciadri}]{Raga02AA}
Raga, A.~C., Vel{\'a}zquez, P.~F., Cant{\'o}, J., \& Masciadri, E.
  2002{\natexlab{b}}, A\&A, 395, 647

\bibitem[{Rubini {et~al.}(2007)Rubini, Lorusso, {Del Zanna}, \&
  Bacciotti}]{Rubini07}
Rubini, F., Lorusso, S., {Del Zanna}, L., \& Bacciotti, F. 2007, A\&A, 472, 855

\bibitem[{Stone \& Hardee(2000)}]{Stone00}
Stone, J.~M. \& Hardee, P.~E. 2000, ApJ, 540, 192

\bibitem[{Stone \& Norman(1993)}]{Stone93}
Stone, J.~M. \& Norman, M.~L. 1993, ApJ, 413, 198

\bibitem[{Te{\c s}ileanu {et~al.}(2012)Te{\c s}ileanu, Mignone, Massaglia, \&
  Bacciotti}]{Tesileanu12}
Te{\c s}ileanu, O., Mignone, A., Massaglia, S., \& Bacciotti, F. 2012, ApJ,
  746, 96

\bibitem[{Wilson(1984)}]{Wilson84}
Wilson, M.~J. 1984, MNRAS, 209, 923

\bibitem[{Yirak {et~al.}(2010)Yirak, Frank, \& Cunningham}]{Yirak10}
Yirak, K., Frank, A., \& Cunningham, A.~J. 2010, ApJ, 722, 412

\bibitem[{Yirak {et~al.}(2009)Yirak, Frank, Cunningham, \& Mitran}]{Yirak09}
Yirak, K., Frank, A., Cunningham, A.~J., \& Mitran, S. 2009, ApJ, 695, 999

\bibitem[{Zinnecker {et~al.}(1998)Zinnecker, McCaughrean, \&
  Rayner}]{Zinnecker98}
Zinnecker, H., McCaughrean, M.~J., \& Rayner, J.~T. 1998, Nature, 394, 862

\end{thebibliography}
\end{document}